\documentclass[12pt]{article}
\usepackage{graphicx}
\newcommand{\lsim}{\lower.7ex\hbox{$\;\stackrel{\textstyle<}{\sim}\;$}}

\begin{document}

\title{
\begin{flushright}
{\rm {\normalsize{MCTP-06-13}}}\\
\vspace{1cm}
\end{flushright}
Adjoint Chiral Supermultiplets and Their Phenomenology}

\author{Yanou Cui
\\[.3cm]
{\small{\it Michigan Center For Theoretical Physics (MCTP)}}\\
{\small{\it Department of Physics, University of Michigan}}\\
             {\small {\it Ann Arbor,MI 48109, USA}} \\
}

\maketitle
\begin{abstract}

\indent

Matter fields in the MSSM are chiral supermultiplets in
fundamental (or singlet) representations of the standard model
gauge group. In this paper we introduce chiral superfields in the
adjoint representation of $SU(3)_C$ and study the effective field
theory and phenomenology of them. These states are well motivated
by intersecting D-brane models in which additional massless
adjoint chiral supermultiplets appear generically in the low
energy spectrum. Although it has been pointed out that the
existence of these additional fields may make it difficult to
obtain asymptotic freedom, we demonstrate that this consideration
does not rule out the existence of adjoints. The QCD gauge
coupling can be perturbative up to a sufficiently high scale, and
therefore a perturbative description for a D-brane model is valid.
The full supersymmetric and soft SUSY breaking Lagrangians and the
resulting renormalization group equations are given.
Phenomenological aspects of the adjoint matter are also studied,
including the decay and production processes. The similarity in
gauge interaction between the adjoint fermion and gluino
facilitates our study on these aspects. It is found that these
adjoint multiplets can give detectable signals at colliders and
satisfy the constraints from cosmology.

\vspace{.5cm}

\end{abstract}

\setcounter{equation}{0}

\section{Introduction}
\indent

     It is well known that in both of the two benchmark models of particle
  physics --- the Standard Model (SM) and its supersymmetric
  extension, the Minimal Supersymmetric Standard Model (MSSM) --- all the matter fields are in the fundamental (or singlet) representations
 under the SM group $SU(3)_C\times SU(2)_L\times U(1)_Y$, and all
 the gauge fields are in the adjoint representation under the
 corresponding gauge group. However, it is interesting to ask whether it is possible to consistently include
 matter fields in higher representations under the gauge groups. This is not an unimportant question to ask since this possibility is also strongly motivated by intersecting
 D-brane models. Intersecting D-brane models are a class of phenomenologically interesting
 string models where the gauge fields of the SM are localized on
 D-branes wrapping certain cycles on an underlying compact
 geometry. Chiral matter emerges at the intersections of
 these D-branes and the family number is determined by the intersection number of the branes.
 However, in addition to the ordinary chiral matter localized at
 intersections, these models typically possess non-chiral open
 string states resulting from the deformation and Wilson line
 moduli of the D-branes\cite{blumenhagen, cvetic}. These states appear as adjoint matter in the four dimensional low-energy effective field theory. Although they have not been observed
 yet, they could be phenomenologically interesting at future colliders,
 such as the LHC, and bring new knowledge to our understanding of elementary
 physics. Especially, an interesting point is: if the adjoint matter originating from D-brane moduli is observed,
 it is possible to be an indirect signal and probe of extra dimensions. As we mentioned, the adjoints in D-brane models
 geometrically originate from the transverse fluctuations of D-branes in higher dimensions. On the other hand, if we look at the T-dual picture,
  these geometric moduli become the higher dimensional components
  of the relevant gauge field, which appear as scalars in the four-dimensional
  theory. Taking into account that these adjoints have gauge couplings to SM fields, once observed, they
  might produce stronger signal at colliders, compared to the KK modes of the graviton as a probe of extra
  dimensions. Of course, adjoint matter fields in four-dimensional
  effective theory can have other sources and may not be relevant
  to extra dimensions in these cases. In any case now we can expect some nontrivial phenomenological implications of the adjoint matter. However, it was pointed out that these
 exotic multiplets should be got rid of in some way since they give additional positive contributions to the gauge coupling
 $\beta$-function. The main motivation for this argument
 was to obtain an asymptotically free gauge theory (for $SU(3)_C$) which
 facilitates perturbative gauge coupling unification enjoyed by
 the MSSM. This consideration has been taken seriously and D-brane models
 without additional adjoint fields have been
 developed\cite{blumenhagen, dudas}. In these specific models, the
 D-branes wrap rigid cycles so that the open string moduli
 are frozen.\\

    However, the possible trouble with asymptotic freedom brought
in by the adjoint fields does not actually impose a no-go theorem
on the existence of new fields in the low energy spectrum. First
of all, unlike in heterotic string models, gauge coupling
unification, as the main motivation for asymptotic freedom, is not
a prerequisite in intersecting D-brane models. Instead, all that
is required is that the gauge coupling be perturbative up to the
string scale. After deriving the RG equations of the gauge
couplings in our model in section 3, we will demonstrate in detail
that generically these additional adjoint fields do respect this
requirement. Furthermore, it is possible to preserve gauge
coupling unification in an intersecting D-brane model with these
new fields. This issue has been discussed in detail in
ref.\cite{lust}, in which the authors demonstrate that under a few
reasonable assumptions, gauge couping unification can be realized
in realistic three generation supersymmetric intersecting brane
models and the result is still valid even when the contributions
from additional non-chiral matter are included. An earlier study of unification with adjoints in SUSY scenario can be found in \cite{jones, jones2}.\\

   Based on the discussions above, we see that adjoint matter
fields are worthy of serious study despite some possible
challenges of phenomenology. In the following sections of this
paper, we will study the effective field theory of these adjoint
matter fields by embedding them into the
MSSM.\footnote[1]{Although inspired by D-brane models, we do not
dwell on string model building in this paper. The considerations
will be based on effective field theory. However, it is plausible
to realize this MSSM+ $SU(3)_C$ adjoints model from a real
intersecting D-brane model, without auxiliary matter fields, e.g.
$SU(2)$ adjoints, symmetric or antisymmetric
matter\cite{blumenhagen},\cite{lust}-\cite{march2}.}

\section{Lagrangian of Chiral  Adjoint Matter}
\indent

    In order to derive the relevant Lagrangian, we need to briefly review the transformation
properties of adjoint matter fields under gauge and supersymmetry
transformations. In the intersecting D-brane models which
reproduce the SM gauge group $SU(3)_C\times SU(2)_L\times U(1)_Y$,
there is an adjoint field for each of the three subgroups. These
fields are in the adjoint representation of the corresponding
subgroup and are in the singlet representation under the other two
subgroups. For each gauge group, the multiplicity $n$ of the
adjoint (an analogy to the family number of ordinary matter) is
determined by the specific configuration of D-branes. Under
supersymmetry, all of the adjoints are in chiral supermultiplets,
as in the case for ordinary matter. Considering the theory above
the electroweak symmetry breaking (EWSB) scale, we require the
Lagrangian to be invariant under the SM gauge group. We will focus
on $SU(3)_C$ adjoints. It can be expected that compared to
ordinary matter, the distinct and interesting features of adjoint
matter come from its distinct representation under the
gauge group.\\

    The complete renormalizable supersymmetric and gauge invariant Lagrangian is:\\
\begin{equation}
\mathcal{L}_1=\mathcal{L}_{g_0}+\mathcal{L}_{g}+\mathcal{L}_{\tilde{\mu}}+\mathcal{L}_{Y}
\end{equation}
 $\mathcal{L}_{g_0}$ is the
kinetic term for free gauge fields and has the usual form as we
see in the MSSM. $\mathcal{L}_{g}$ is the connective gauge
interaction between the adjoints and the gauge fields ($SU(3)_C$
in this paper). $\mathcal{L}_{\tilde{\mu}}$ is the mass term of
the adjoints. The non-chiral property of the adjoints allows this
gauge invariant mass term. The fermionic component of the adjoint
superfield is Majorana which is an important point for our
discussion on phenomenological issues later. Although these
adjoints typically get mass only after supersymmetry breaking,
SUSY breaking effects in the K\"{a}hler potential in supergravity
can result in such a supersymmetric mass term from $F^\dag$
insertions, which is similar to the Giudice-Masiero mechanism in
the MSSM\cite{giudice}. $\mathcal{L}_{Y}$ is the Yukawa
interaction between the adjoints.\\

  Let us now address each term of the Lagrangian in turn in terms
of the F-term and D-term of the relevant chiral superfield and
vector superfield. We will mainly use the notation and conventions of refs.\cite{wess, drees}.\\
\begin{eqnarray}
\mathcal{L}_1&=&\left(\int d^2\theta\frac{1}{16kg^2}\tilde{Tr}
W^{\alpha}W_{\alpha}+h.c.\right) +\int d^4\theta
\frac{\tilde{Tr}(\Phi^{\dag a}t^ae^{2gV^cT^c}\Phi^a
t^a)}{l}\\\nonumber &+&\left[\int
d^2\theta(\frac{1}{2}\tilde{\mu}\frac{\tilde{Tr}(\Phi^{a}t^a\Phi^
{b}t^b)}{l}+\tilde{y}\frac{\tilde{Tr}(\Phi^{a}t^a\Phi^
{b}t^b\Phi^{c}t^c)}{l^{3/2}})+h.c.\right]
\end{eqnarray}
where $\Phi^at^a$ is the adjoint chiral superfield with color
index
$a$ and can be expanded in component fields as\\
\begin{equation}
\Phi=A+\sqrt{2}\psi\theta+F\theta\theta
\end{equation}
$t^a$ is $N\times N$ matrix for $SU(N)$ group where $N=3$ in this
case. $V$ is the usual vector superfield, whose matrix $T^c$ in
this case is $8\times8$ when acting on adjoint matter.
$\tilde{\mu}$ is the supersymmetric Majorana mass of the adjoint
field. $y$ is the Yukawa coupling constant. $k$ is the Dynkin
index of the representation of the gauge field according to the
matter field it acts on. $l$ is defined by
\begin{equation}
Tr(t^at^b)=l\delta^{ab}
\end{equation} \\
For the fundamental representation, $l=\frac{1}{2}$.

   The combinations of color indices $(a,b)$ and $(a,b,c)$ in eq.(2) run over all the possible
combinations. The notation of $\tilde{Tr}$ means the symmetrized
trace of the invoved matrices which gives the invariant tensor of
certain representation. This notation is no different from the
usual $Tr$ if two matrices are included. When three matrices are
involved, this new notation sums over all the possible invariants.
$\tilde{Tr}(t^at^bt^c)$ is actually exactly the form of triangle
anomaly which we are familiar with. An interesting upshot here is
that there is no gauge invariant Yukawa interaction term for
$SU(2)$ adjoint since the triangle anomaly for an $SU(N)$ group is
non-vanishing only for $N>2$. The following results are useful for
our later discussion in this paper\cite{math}:
\begin{eqnarray}
Tr(\{t_a,t_b\}t_c)=\frac{1}{2}d_{abc}\\
\sum_{b,c}d_{abc}d_{dbc}=\frac{N^2-4}{N}\delta_{ad}\\
\sum_{a,b,c}d_{abc}d_{abc}=\frac{(N^2-4)(N^2-1)}{N}\\
Tr([t_a,t_b]t_c)=i\frac{1}{2}f_{abc}\\
\sum_{c,d}f_{acd}f_{bcd}=N\delta_{ab}
\end{eqnarray}

  We would like to point out some important properties of the adjoint
fields. From eq.(2), we see that in the gauge invariant
renormalizable Lagrangian, they only couple to $SU(3)_C$ gauge
fields (including gluons and gluinos) of the MSSM. They can only
directly couple to Higgs and quarks through non-renormalizable
interactions which we will consider in section (2.2). For a
complete MSSM augmented by chiral adjoint matter, it is fair to
include the connective gauge terms for ordinary matter and the
Higgs Yukawa terms. But they are simply in the usual form we know
well. These terms will contribute to the RG equation of the gauge
coupling, but not to the
1-loop RG running of adjoint masses. \\

  In order to obtain RG equations and discuss phenomenological
issues, we derive the explicit form of the Lagrangian by component
fields:
\begin{eqnarray}
\mathcal{L}_{\tilde{\mu}}&=&\tilde{\mu}(A^aF^a-\psi^a\psi^a)+h.c.\\
\mathcal{L}_{g0}&=&-\frac{1}{4}v^{c}_{\mu\nu}v^{c}_{\mu\nu}+i\bar{\lambda}^{c}\bar{\sigma}^{\mu}{\mathcal
D}_{\mu}\lambda^{c}+\frac{1}{2}D^{c}D^{c}\\
\mathcal{L}_Y&=&\sqrt{2}\tilde{y}d^{abc}(F^aA^bA^c-\psi^a\psi^bA^c)+h.c
\end{eqnarray}
\begin{eqnarray}
\mathcal{L}_g&=&\frac{Tr(t^at^a)}{l}[{\mathcal
D}_{\mu}A^{a\dag}{\mathcal
D}^{\mu}A^a+i\bar{\psi}^a\bar{\sigma}^{\mu}{\mathcal
D}_{\mu}\psi^a+F^{\dag
a}F^a-\sqrt{2}g(\bar{\lambda}^{c}\bar{\psi}^aT^cA^a+h.c.)\\\nonumber
&&+gA^{\dag a}D^cT^{c}A^a]\\\nonumber &=&(\partial_{\mu}A^{\dag
a}\partial^{\mu}A^{a} +g^2f^{acd}f^{aef}v^c_{\mu}v^{e\mu}A^{\dag
d}A^f-gf^{acd}(v^c_{\mu}A^{\dag
d}\partial^{\mu}A^a+v^c_{\mu}\partial^{\mu}A^{\dag
a}A^d)\\\nonumber
&&+(i\bar{\psi}^a\bar{\sigma}^{\mu}\partial_{\mu}\psi^a-igf^{acd}\bar{\psi}^a\bar{\sigma}^{\mu}v^c_{\mu}\psi^d)
+F^{\dag a}F^a+i\sqrt{2}gf^{cad}(-A^{\dag
d}\lambda^c\psi^a+\bar{\psi}^a\bar{\lambda}^cA^d)\\\nonumber
&&-igf^{cad}A^{\dag a}D^cA^d
\end{eqnarray}
where $v_{\mu}$ is the gauge boson, $\lambda$ is gaugino, $D$ is
the auxiliary field
\begin{equation}
 v^{a}_{\mu\nu}=\partial_{\mu}v^{a}_{\nu}-\partial_{\nu}v^{a}_{\mu}-gf^{abc}v^{b}_{\mu}v^{c}_{\nu}
\end{equation}
\begin{equation}
{\mathcal
D}_{\mu}\lambda^{a}=\partial_{\mu}\lambda^{a}-gf^{abc}v_{\mu}^{a}\lambda^{c}\end{equation}
Eq.(15) is the generic expression for the covariant derivative of
the adjoint representation which applies for both gaugino and the
adjoint fields $\psi$,$A$  which we used in getting
eq.(13)).\\

  We can obtain the explicit expression of the auxiliary fields
  from the equation of motion:
\begin{eqnarray}
F^a&=&-\sqrt{2}\tilde{y}^*A^{\dag b}A^{\dag c}d^{abc}-\tilde{\mu}^* A^{\dag a}\\
F^{\dag a}&=&-\sqrt{2}\tilde{y}A^{b}A^{c}d^{abc}-\tilde{\mu} A^{a}\\
D^c&=&igf^{cad}A^{\dag a}A^d
\end{eqnarray}
\\
  In order to study the low energy phenomenology, in addition to the manifestly supersymmetric Lagrangian $\mathcal{L}_1$,
   we must include soft breaking terms, which we collectively denote as $\mathcal{L}_2$. This is analogous to the usual case\cite{martin}:
\begin{equation}
\mathcal{L}_2=-\frac{1}{2}(M\lambda^a\lambda^a+c.c.)-m^2A^{\dag
a}A^a-\left(\frac{1}{2}bA^aA^a+\sqrt{2}d^{abc}aA^aA^bA^c+c.c.\right)
\end{equation}

   Since the adjoint fields do not couple to ordinary matter directly through renormalizable interactions, we need to include dim-5
non-renormalizable terms suppressed by some high scale $M_*$ in
order to study the phenomenology at colliders. In order to
simplify, we assume that flavor symmetry breaks at a scale above
$M_*$ and that the Yukawa matrix ($Y'$) for such dim-5
interactions has the similar hierarchical structure as the Yukawa
in the SM, since they share the same flavor physics at high scale.
Concretely, this means that all the mixings are negligible and the
only non-vanishing entry of the Yukawa matrices is (3,3), with the
value $y'_t$ or $y'_b$ . If we further assume that the hierarchy
between the top quark mass and bottom quark mass mainly comes from
their difference in Yukawa coupling, $y'_b$ is also negligible.
Under the above assumptions, only $y'_t\sim1$ enters our
discussion. (Of course, if we need, it is not hard to write down
the full dim-5 Lagrangian without those assumptions. However, the
expressions will be more complicated.) Here we use $T_a$ instead
of $t^a$ to represent the matrix representation of the adjoint
field, in order not to mix up with the notation $t^c$ for the
conjugate of
right-handed top quark field. Now we get:\\

\begin{eqnarray}
\mathcal{L}_Y^{'}&=&\int
d^2\theta\frac{1}{M_*}Q_3\Phi_aT_at^cH_u+h.c.\\\nonumber &=&\int
d^2\theta\frac{1}{M_*}\Phi_aT_a(t^ctH^0_u-t^cbH^+_u)+h.c.\\\nonumber
&=&\frac{1}{M_*}T_a[A_a(F_tH^0_u\tilde{t^c}+F_{H_u^0}\tilde{t}\tilde{t^c}+F_{t^c}H^0_u\tilde{t}-t^ctH_u^0-t^c\tilde{H_u^0}\tilde{t}
-t^ctH_u^0)\\\nonumber
&+&2\psi_a(t^cH^0_u\tilde{t}+\tilde{t^c}\tilde{H_u^0}\tilde{t}+t\tilde{t^c}H^0_u)+F_a\tilde{t^c}H_u^0\tilde{t}-(t\rightarrow
b, H^0_u\rightarrow H^+_u)]+h.c.
\end{eqnarray}

\section{Renormalization Group Equations (RGEs) and Asymptotic Freedom Problem}
\subsection{RGEs}
\indent

  In order to study phenomenology at the low scale (electroweak scale, typically), we need to derive
RGEs to see how the parameters evolve. Our results are obtained
from explicit diagrammatic techniques, based on conventional
Feynman diagrams. According to the Lagrangian obtained in section
2, using $\overline{\rm DR}$\cite{buras} , we can obtain RGEs for all the relevant parameters of the theory.\\

  First let's look at the 1-loop RGE of scalar mass squared $\tilde{m}^2=m^2+|\tilde{\mu}|^2$. Note that
in this model this quantity is the sum over soft SUSY breaking
$\tilde{m}^2$ and the `supersymmetric' $|\tilde{\mu}|^2$ which
results from the non-chiral property of the adjoints. For
simplicity, we can assume $\tilde{\mu}$ is real so that it is
rightly the physical mass of $\Psi$. $M$ is the gaugino mass.
$t=\ln(p/p_0)$, where $p$ is the renormalization scale, $p_0$ is
an arbitrary reference scale.
\begin{equation}
\frac{d\tilde{m}^2}{dt}=\frac{2}{(4\pi)^2}\left\{-g^2\left[12(M^2+\tilde{\mu}^2)-9\tilde{m}^2\right]+|\tilde{y}|^2(10\tilde{m}^2-\frac{40}{3}\tilde{\mu}^2)+\frac{10}{3}|a|^2\right\}
\end{equation}
1-loop RGE of b-term is
\begin{equation}
\frac{db}{dt}=\frac{2}{(4\pi)^2}\left[\tilde{\mu}\left(24g^2M+\frac{20}{3}\tilde{y}^*a\right)+b\left(9g^2+\frac{10}{3}|\tilde{y}|^2\right)\right]
\end{equation}
   The physical squared masses $x^2$ of the scalar adjoint are
\begin{equation}
x^2=\tilde{m}^2\pm|b|
\end{equation}
\\
    The fermion mass does not get enhanced from soft SUSY breaking
terms. The 1-loop RGE of the fermion mass $\tilde{\mu}$ is
\begin{equation}
\frac{d\tilde{\mu}}{dt}=\frac{2}{(4\pi)^2}
\left(\frac{3}{4}g^2+\frac{5}{3}|\tilde{y}|^2\right)\tilde{\mu}
\end{equation}

  Based on these RGEs, we are ready to analyze the evolution of these mass parameters. When running towards lower scale, the
fermion mass $\tilde{\mu}$ is strictly reduced by gauge coupling
and Yukawa coupling. The situation is not so definite for
$\tilde{m}^2$ and $b$, depending on the evolutions and relative
sizes of different soft parameters. Nevertheless we can draw the
conclusion that the physical squared mass $x^2$ does not go
tachyonic in a large region of parameter space, and it is
reasonable to assume that at the EW scale, the mass of the scalar
adjoint $A$ is larger than that of the fermionic adjoint $\psi$.
This case is very similar to that of the MSSM: all the scalar
fields are heavier than their fermionic partners because of the
enhancement from soft SUSY breaking terms. In section 4, we will
take this case as a typical example to study the decay
process of adjoint fields at colliders.\\

  Now let's examine the evolution of the $SU(3)_C$ gauge coupling $g$. Since we already know the matter content of the model, we can
directly apply the general RGEs \cite{martin2}. We give the
results up to 2-loop contribution. Our important discussion about
asymptotic freedom problem will be based on these.
\begin{eqnarray}
\frac{dg}{dt}&=&\frac{1}{16\pi^2}\beta_g^{(1)}+\frac{1}{(16\pi^2)^2}\beta_g^{(2)}\\\nonumber
\beta_g&=&\beta_{g MSSM}+\beta_{g \Phi}\\
\beta^{(1)}_{g MSSM}&=&-3g^3\\
\beta^{(2)}_{g MSSM}&=&g^3\left(\frac{11}{5}g_1^2+9g^2_2+14g^2-4|y_t|^2\right)\\
\beta_{g \Phi}^{(1)}&=&3ng^3\\
\beta_{g \Phi}^{(2)}&=&54ng^5-\frac{9}{4}ng^3|\tilde{y}|^2
\end{eqnarray}
Summing over eq.(26)-(29), taking them into eq.(25), we get the
full RGE for $g$ up to 2-loop corrections:
\begin{equation}
\frac{dg}{dt}=\frac{1}{16\pi^2}g^3(-3+3n)+\frac{1}{(16\pi^2)^2}g^3\left[\frac{11}{5}g_1^2+9g^2_2+(14+54n)g^2-4|y_t|^2-\frac{9}{4}n|\tilde{y}|^2\right]
\end{equation}
  In the 2-loop contribution from the MSSM
eq.(27), for simplicity we assume that the only Yukawa that is
considerable is $y_t\sim1$. Eq.(28) and (29) are additional
contributions from the adjoint fields. $n$ is the family number of
the adjoint fields
which is determined by the D-brane configuration.\\

  Before going to the discussion of asymptotic freedom using eq.(30), we write down the 1-loop RGE of gaugino mass\cite{martin2}:
\begin{equation}
\frac{dM}{dt}=\frac{1}{8\pi^2}\frac{\beta_g}{g}M
\end{equation}

\subsection{Adjoint Matter Fields and Asymptotic Freedom}
\indent

  As we know, asymptotic freedom behavior of the strong coupling enables gauge coupling unification, or at least helps preserve
the perturbative property of the gauge coupling at high scale. We
also know that adding any new matter fields to the MSSM will make
it more difficult to preserve the asymptotic freedom property.
This is why many suggest getting rid of the adjoints emerging from
intersecting D-brane models, although these states are too generic
to dismiss immediately. As we mentioned in the introduction: gauge
coupling unification is not a prerequisite for intersecting
D-brane models and the corresponding requirement in these models
should be that the gauge coupling does not diverge below the
string scale, which can comfortably be the intermediate scale.
This means that for D-brane models, as long as the theory is still
perturbative at high scale, asymptotic
freedom is unnecessary.\\

  First let's see if it is still possible to realize asymptotic freedom with these
additional adjoint fields. From eq.(30), we see that when $n>1$,
the 1-loop $\beta_g$ is positive. From the dominance of the 1-loop
correction over the 2-loop correction, we can draw the conclusion
that asymptotic freedom cannot be realized for $n>1$. When $n=1$,
it is interesting that the 1-loop $\beta$ function is $0$. So we
need to look into the 2-loop correction in this case.
Unfortunately, it is easy to see that 2-loop contribution is
positive unless the Yukawa coupling $y$ is unnaturally large.
However, the divergence scale in this case can hopefully be above
string scale (preferably
 some intermediate scale), especially when threshold corrections are taken into account.
  We will demonstrate how threshold correction can lift the divergence scale up to string scale for $n>1$ cases.
  Similarly, we can expect the divergence scale to be as high as the string scale for the case of $n=1$ since the divergence rate of 2-loop correction is $\frac{1}{16\pi^2}$ suppressed, compared to that of 1-loop correction.\\

    Now we see it is truly hard to preserve asymptotic freedom for MSSM+chiral adjoint model. We need to analyze
the divergence scale of the gauge coupling, which is preferred to
be above the string scale in order to preserve the validity of
perturbation theory. The solution of the 1-loop RGE for $g$ is:
\begin{equation}
g^2(p)=\frac{g^2(p_0)}{1+\frac{g^2(p_0)}{(4\pi)^2}(-3n+3)\ln(\frac{p^2}{p^2_0})}
\end{equation}
where $p_0$ is the reference scale. \\

We can solve for the divergence scale $\Lambda$ from eq.(32):
\begin{equation}
\Lambda=p_0\exp\left[\frac{2\pi}{\alpha_3(p_0)}\frac{1}{3n-3}\right]
\end{equation}
where as usual, $\alpha_3=\frac{g^2}{4\pi}$.\\

  First we assume the threshold where all the MSSM particles and the $SU(3)_C$
adjoints enter the loop diagrams is just above the scale $m_Z$ (we
actually assume a degenerate SSM spectrum for simplicity). Then
$p_0=m_Z=91$ GeV. According to experimental result,
$\alpha_3(m_Z)=0.119\pm0.002$. Applying this to eq.(33), we find:
\footnote[2]{In the relevant D-brane models, the family number of
adjoints can be up to 3. That is why we consider the $n=1,2,3$
cases}\\
for $n=1$, $\Lambda=\infty$\\
for $n=2$, $\Lambda\sim 4\times10^9$ GeV\\
for $n=3$, $\Lambda\sim 6\times10^5$ GeV\\

  Therefore, to 1-loop order when $n=1$, the strong coupling in our
new model is always perturbative. When $n=2$, $g$ diverges a bit
lower than the intermediate scale which seems not so good if we
assume the string scale to be there. However, if we assume the
string scale is at the TeV scale as is the case in large extra
dimension scenario, even the $n=2$ and $n=3$ cases are
satisfactory.
\\

  Furthermore, it is too restrictive to assume that all
beyond-the-SM particles run in the loops just above $m_Z$. Let's
assume the SM RGE dominates (where $\beta^{(1)}=-7g^3$) until the
threshold at  $1$ TeV where MSSM particles and the adjoints run
in. In this case, we get
more satisfactory results:\\
for $n=1$, $\Lambda=\infty$\\
for $n=2$, $\Lambda\sim 1.2\times10^{13}$ GeV\\
for $n=3$, $\Lambda\sim 1.1\times10^8$ GeV\\
Now for $n=2$, $g$ diverges well above intermediate scale.\\

  Thus, we find that although it is hard to preserve asymptotic freedom in our model,
when $n$ is small it is still possible to keep the theory
perturbative up to the string scale (either intermediate or low
scale) in the intersecting D-brane scenario and to do so becomes
even easier when that scale is low. It is also possible that
threshold corrections can further lift the divergence scale.
Especially, when $n=1$, $g$ is always perturbative at any high
scale (to 1-loop order). These solidify the motivation of this
effective field theory model with chiral adjoint fields.
Furthermore, it should be kept in mind that in more general
intersecting D-brane models which contain other beyond-the-MSSM
particles in addition to the adjoint matter addressed in this
work, it may be possible to further increase the divergence scale
or even to restore gauge coupling unification\cite{lust} (however,
see\cite{berenstein}).

\section{Phenomenology of Chiral Adjoint Fields}
\indent

  We are now ready to study the phenomenology of these $SU(3)_C$ chiral adjoint
states: their behavior at colliders and the role they play in
cosmology. As mentioned in section 3.1, in the following
discussion, we will assume the mass of the scalar component $A$ is
larger than that of the fermionic component $\psi$. Based on this
assumption, whether $\psi$ is the LSP or not (the lightest
particle in the MSSM+adjoints model) will result in different
stories, which we will discuss respectively.
\subsection{Decays}
\indent

  The lifetime or decay process of a particle is an important issue to study in phenomenology since it is relevant to both its
cosmological ramifications (whether satisfies the constraint from
BBN) and collider physics (whether decays promptly enough to be
detected). Now let's study this issue for $A$$(A^\dag)$  and
$\psi$ in
turn.\\

  The possible decay of $A$$(A^\dag)$ comes from the
gauge interaction
\begin{equation}
  L_{dec1}=i\sqrt{2}gf^{cad}(-A^{\dag
d}\lambda^c\psi^a+\bar{\psi}^a\bar{\lambda}^cA^d)
  \end{equation}
  We also assume that
  $m_A>m_{\psi}+m_\lambda$ to allow an on-shell two-body decay. We
  will calculate the decay rate of the process
  $A^{\dag}\rightarrow\Psi+\Lambda$ as an example. Here we have constructed the 4-component Majorana particles $\Psi$ and
  $\Lambda$ from the original Weyl fermions.\\

  If $m_{A^{\dag}}\gg m_\Psi,m_\Lambda$ the
integration over phase space simplifies, and the decay rate
is\cite{gordy}:
\begin{equation}
\Gamma_1=\frac{3}{8\pi}g^2m_{A^\dag}
\end{equation}
We know that for the SU(3) gauge coupling,
$\frac{g^2}{4\pi}\simeq0.1$. To make a numerical estimation, we
assume that $m_\Psi,m_\Lambda\sim100$ GeV,$m_{A^\dag}\sim1$ TeV.
Then taking the numbers in eq.(35):
\begin{equation}
\Gamma_1\simeq150\, {\rm GeV}\simeq2\times10^{26}\, {\rm sec}^{-1}
\end{equation}
We see that this decay is a prompt process and therefore can be
detected at a collider. Meanwhile, because the lifetime is much
shorter than 1 sec, $A^\dag$ ($A$) will not violate the BBN
constraints from cosmology.\\

      Now we come to study decay processes of $\Psi$. If $\Psi$ is
   the LSP or very long-lived, we need to see if it
   respects the cosmological constraints, which we will discuss in the
   third subsection. If $\Psi$ is not the LSP, three-body on-shell decays are
   possible according to the nonrenormalizable term in the
   Lagrangian:
\begin{equation}
\mathcal{L}_Y^{'}=\frac{1}{M^*}T_a[2\psi_a(t^cH^0_u\tilde{t}+\tilde{t^c}\tilde{H_u^0}\tilde{t}+t\tilde{t^c}H^0_u)-(t\rightarrow
b, H^0_u\rightarrow H^+_u)]+h.c.
\end{equation}

 To give an example of the decay rate, we will focus on the second
 term, which results in a $4t+E\!\!\!\!/$ signature (Fig.1) at colliders.
 With little background from SM, and with sufficient signal cross section, this signature can be very
 interesting evidence for new physics. We will find that this
 is quite plausible at the end of section 4.3 after we make an estimation of the pair
 production cross section of $\Psi$. Here we focus on giving the
 decay rate of the typical process:
 $\Psi\rightarrow\chi_1+\tilde{t}_1+\tilde{t}_1^c$, where we
 transformed $\psi$ and $\tilde{H}_u$ from Weyl basis
to Majorana basis and rotated from $SU(2)$ eigenstates to mass
eigenstates. If we assume $\chi_1$ is the LSP, then the outgoing
$\chi_1$ results in pure missing energy signature. The rotation
from gauge eigenstates to mass eigenstates results in the factors
$\frac{1}{2}\sin2\theta_{\tilde{t}}$ and $N_{31}$. Again,to
simplify the integration over the phase space of this 3-body
decay, we assume that $m_\Psi\gg
m_{\chi_1},m_{\tilde{t}_1},m_{\tilde{t}_1^c}$. We get the decay
rate\cite{gordy}:
\begin{equation}
\Gamma'=\frac{1}{4(4\pi)^3}(\sin2\theta_{\tilde{t}})^2
N_{31}^2\frac{m_\Psi^3}{M_*^2}
\end{equation}
\begin{figure}[h]
\vspace{0cm} \center
\includegraphics[width=8cm]{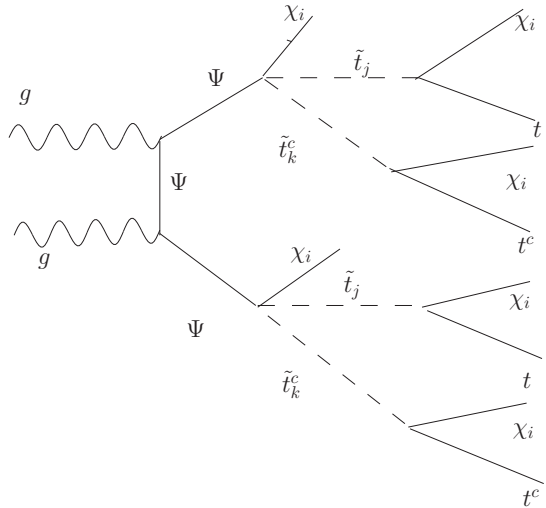}\\
\caption{$4t+E\!\!\!\!/$ signal from our model: g is gluon,
$\chi_i$ is neutralino which result in the signal of missing
energy, i=1,2,3,4. For mass eigenstates of stops, j,k=1,2}
\end{figure}

To make a numerical estimate, we assume the mixing factors to be
of order 1, and $m_\Psi\sim1 $TeV. For the three typical suppression scales, the results are: \\

\begin{tabular}{|c|c|}
  \hline
  Suppression Scale $M_*$& Decay rate/Lifetime of $\Psi$\\
  \hline
   1 TeV (low scale)& $10^{-1}$ GeV$/10^{-23}$ sec \\
  $10^{11}$ GeV(intermediate scale) & $10^{-17}$ GeV/$10^{-7}$ sec \\
  $10^{18}$GeV (Planck scale) & $10^{-31}$ GeV/$10^7$ sec \\
  \hline
\end{tabular}

  We see that for TeV or intermediate scale suppression, the decay of $\Psi$ is prompt and detectable at colliders, and thus will
not violate constraints from BBN. However, for higher suppression
scales, some additional mechanism may need to be invoked for
$\Psi$ decays so as to prevent late $\Psi$ decays from violating the BBN constraints.\\

Again, we will discuss more about the $4t+E\!\!\!\!/$ signal at
the end of the next subsection, which should be detectable at
colliders if the suppression scale is not too high.

\subsection{Pair Production of $\Psi$ at Hadron Colliders}
\indent

 Another interesting aspect of phenomenology is production. Since
 $A$ typically decays promptly once produced and is heavier, we will focus on
 $\Psi$. As we see in the derived renormalizable Lagrangians for the adjoints,
 the only direct coupling between the adjoints and SM fields is
 through the connective term of the gauge interaction. For $\Psi$, we
 see this is
 \begin{equation}
 \mathcal{L}_{\psi-g}=i\bar{\psi}^a\bar{\sigma}^{\mu}{\mathcal
D}_{\mu}\psi^a
 \end{equation}
 Here a striking equivalence resulted from gauge invariance is that this is exactly the same
form with the gluino contribution in the kinetic term of gauge
interaction:
\begin{equation}
 \mathcal{L}_{\lambda-g-1}=i\bar{\lambda}^a\bar{\sigma}^{\mu}{\mathcal
D}_{\mu}\lambda^a
\end{equation}
Since both the gluino and $\psi$ are in the adjoint representation
of $SU(3)$, their covariant derivatives $\mathcal{D}_\mu$ are in
the same form as in eq.(15). Meanwhile, both the gluino and $\psi$
are Majorana, eq.(39) and (40) will still take the same form after
transforming to the Majorana basis. Therefore, the diagrams will
have exactly the same Feynman rules in this form (they can have
quite different mass parameters). Or to say it more accurately,
formally, gluino and $\Psi$ have exactly the same interactions
with gluons. Since there have been much study and existing checks
of the phenomenology of gluinos, we can refer to these references
for $\Psi$ because of their close
similarity\cite{pair2}-\cite{nima}. We will discuss some details
below. However, here it is also important to notice that the major
difference between them is that the gluino can couple directly to
quarks and squarks through a connective term in the gauge
interaction:
\begin{equation}
\mathcal{L}_{\lambda-g-2}=-\sqrt{2}g(\tilde{q}^{+}t^{a}q\lambda^{a}+\bar{\lambda}^{a}t^{a}\tilde{q}\bar{q})
\end{equation}
On the other hand, as the component of a chiral superfield, $\Psi$
does not have this direct renormalizable coupling to quarks and
squarks. This is an important
point we need to keep in mind when considering the $\Psi$ collider phenonmenology.\\

  The important pair production processes at
leading order for gluino\cite{pair} and $\Psi$ are shown in Fig.2
and Fig.3 respectively.
\begin{figure}[h]
\vspace{0cm} \center
\includegraphics[width=12cm]{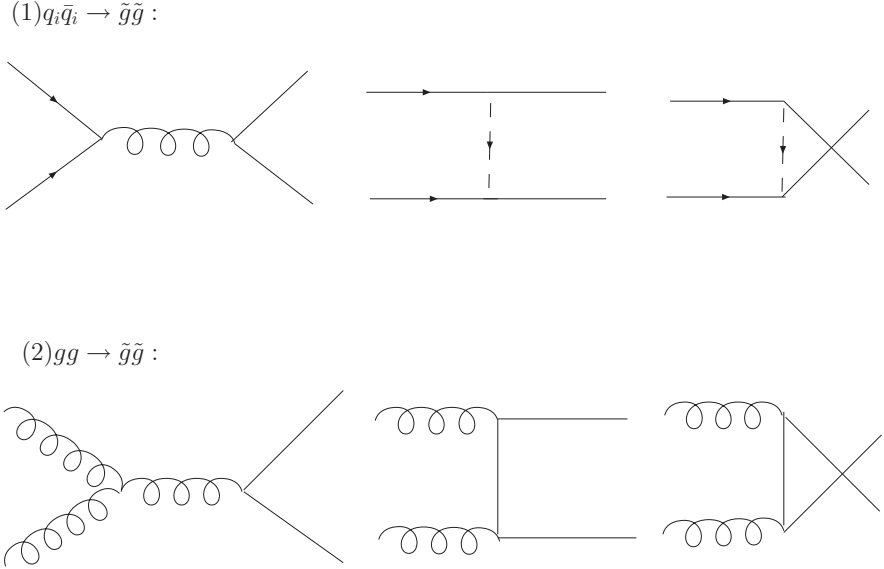}\\
\caption{Pair production of gluino at LHC or other hadron
colliders: the solid line with arrow is quark, solid line without
arrow is gluino, the screw line is gluon, the dot lines are
squarks}
\end{figure}
\begin{figure}[h]
\vspace{0cm} \center
\includegraphics[width=12cm]{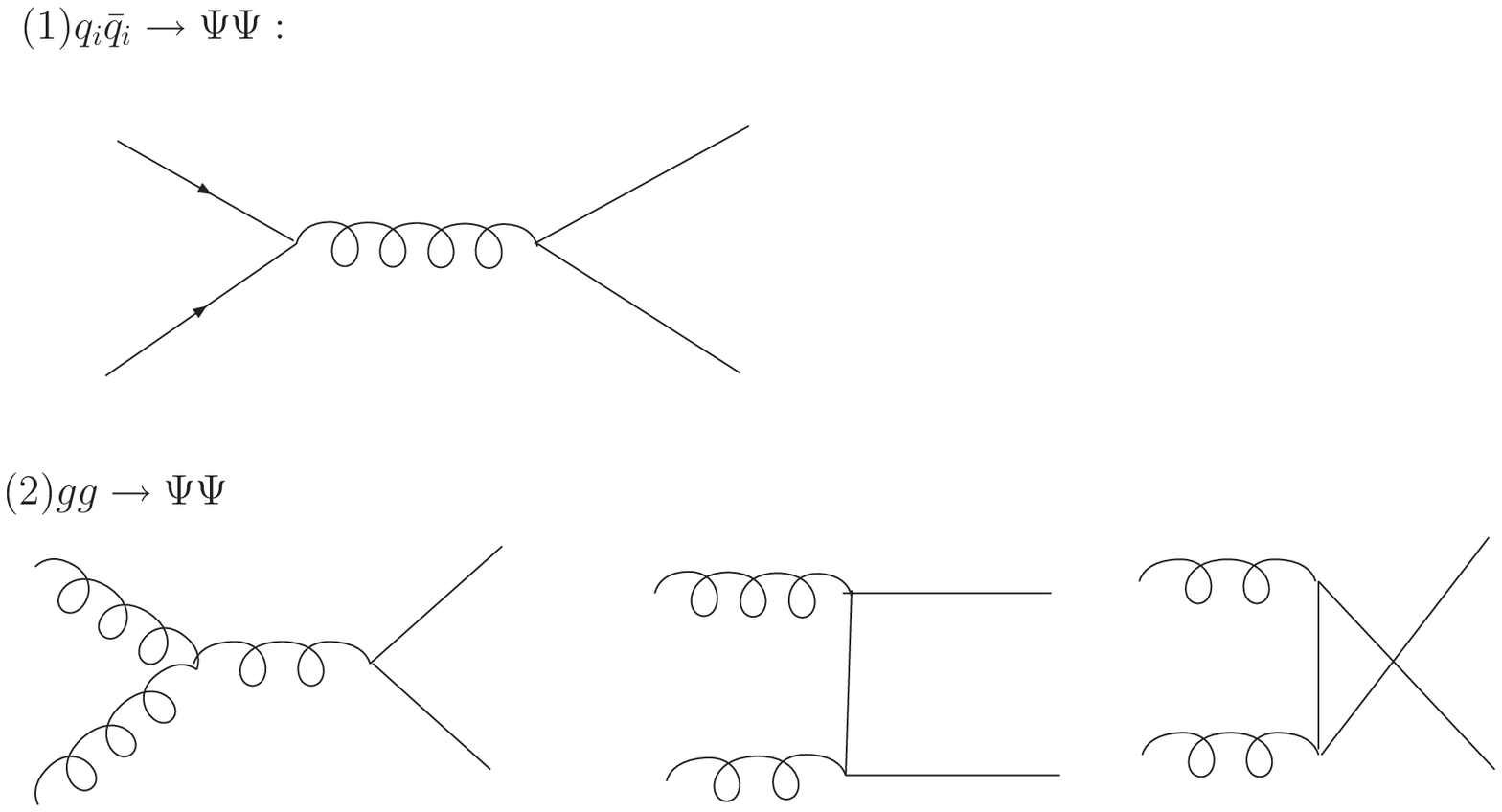}\\
\caption{Pair production of $\Psi$ at LHC or other hadron
colliders: the solid line with arrow is quark, solid line without
arrow is $\Psi$, the screw line is gluon}
\end{figure}
    For production mechanism (1), i.e. produced by
quark-antiquark pair, the gluino has three channels. However, for
$\Psi$, only the first channel of mechanism (1) is allowed by the
Lagrangian (no direct coupling to $q,\bar{q}$). On the other hand,
they all have the same three channels in the production mechanism
(2), i.e. produced by gluon pair. For the gluino, the amplitude of
the total production cross-section is already calculated for both
the Tevatron and the LHC\cite{pair}. Therefore, we can apply these
known results to $\Psi$ to calculate the total production
cross-section of $\Psi$. Numerically, according to their result,
at LHC where$\sqrt{s}$=14TeV, the total cross-section for gluino
pair production ranges from $10^3$pb to $10^{-1}$pb when
$m_{\tilde{g}}$ ranges from 200GeV to 1TeV. If we reasonably
assume that the amplitudes of the two channels involving squarks
are negligible (if the squarks are very heavy) or of the same
order with the other channels, we can expect a very similar
evaluation on production cross-section for our adjoint fermion. In
a word, we can expect the pair production cross-section of $\Psi$
to be large at the LHC. Now we see that the similarity in gauge
interaction between gluino and $\Psi$ facilitates our study on the
collider physics of $\Psi$. \\

   We are also interested in the detection of $\Psi$ once it is produced. If stable, $\Psi$ will also form
R-hadrons, as what happens to long-lived gluinos. This case has
been discussed in refs.\cite{cheung, raby,hewett} for gluinos. If
$\Psi$ is not stable, we need to return to the discussion of its
non-renormalizable decay in section 4.1.2. The 4t signature arises
here, as was mentioned earlier. The first three terms in eq.(37)
can result in $4t+X$ signature at colliders, where $X$ can be
neutralinos (missing energy, if the neutralino is the LSP) or
gluinos. Since it may be hard to discern X, we can make the
inclusive signature of $4t+X$ as a signature (or rather, the many
energetic states from the t decays). Such a signature has very low
background from the SM and therefore can be
 good evidence for beyond the SM physics if it has a reasonable
 (i.e., many fb) cross-section. The two largest sources of 4t background from the SM are: $gg\rightarrow t\bar{t}t\bar{t}$ and QCD
 background. Their total cross-section is less than 10fb (after
 all the cuts, about 1fb) \cite{james}-\cite{4t2}. In order to estimate its cross-section in our
model, we need to know the pair production cross-section of $\Psi$
and the branching ratio of the $4t+X$ signature. As we mentioned,
utilizing analogous calculations of gluino production\cite{pair},
the pair production cross-section of $\Psi$ can be of the order of
pb. Taking into account all the terms in eq.(37), we know that all
the possible decay products of the $\Psi$ pair are: $4t+X_1$,
$4b+X_2$, $2b+2t+X_3$. If we assume that the mass of $\Psi$ is
much larger than the masses of its decay products, different decay
channels have negligible difference in the integration over phase
space and therefore have almost the same branching-ratio. In
conclusion, the branching ratio of $4t+X$ is about $\frac{1}{4}$,
and therefore its cross-section can be around pb as long as
$m_{\Psi}\lsim2$ TeV. In this way, compared with the fb background
from the SM, $4t+X$ can be a notable signature at hadron
colliders.
\subsection{Cosmology of $\Psi$}
\indent

   Based on the discussion of the non-renormalizable decay of $\Psi$,
we see that the lifetime of this adjoint fermion can span a wide
range, depending on the suppression scale $M_*$. For a TeV scale
$\Psi$, it is safe cosmologically if $M_*$ is lower or not much
higher than the intermediate scale. Otherwise we need to check if
it respects the constraints from BBN. If it is very long-lived or
is the LSP, it can be a cold dark matter (CDM) candidate and we
need to check if its relic abundance does not overclose the
universe. An even stronger constraint, which applies to long lived
particles charged under $SU(3)_C$ (including stable gluinos) comes
from the fact that any such particles will bind into nuclei to
produce anomalously heavy elements. However, current experiments
put severe limits on the abundance of heavy elements on the earth
today\cite{smith}. The limits range from $\eta<3.5\times10^{-30}$
for $M=100$ GeV to $\eta<8\times10^{-28}$ for $M=1.2$ TeV, where
$M$ is the mass of the anomalously heavy hydrogen, composing the
anomalous water molecules. Here, $\eta$ is the `anomalous
concentration', namely the number of anomalous water molecules per
usual water molecule in the oceans. With such a small relic
density, it appears unlikely for $\Psi$ to be a dark matter
candidate. On the other hand, this severe bound can be weakened if
we assume that the heavy elements sink towards the center of the
earth so that they have not been discovered yet in the experiments
on the ground\cite{pierce}. Furthermore, we should notice the fact
that in these known experiments that give severe bound on the
existence of stable charged particles, the samples are taken from
terrestrial water, which may not reflect the true abundance of
heavy elements in the whole universe\cite{glashow}. If these are
true, we can still speculate that $\Psi$ is a CDM candidate if it
is stable. Now that we have outlined the constraints on $\Psi$
from cosmology, we would
like to see what they imply for our model.\\

    For a $\Psi$ of intermediate lifetime,
the constraints from BBN are relevant. Once again, we can take
advantage of the analogy to gluino physics. There has been careful
study about the gluinos on this aspect. A typical result
\cite{pierce} is that in split supersymmetry, a TeV mass gluino
must have a lifetime shorter than 100 seconds in order not to
change the abundances of D and $^6$Li. However, the corresponding
study of $\Psi$ can be somewhat different from that of the gluino
since they have different decay processes, which are relevant for
the BBN constraint. Of course, if the relic density of $\Psi$ is
already small enough during the BBN era, its influence on BBN is
negligible. \\

    For a very long-lived $\Psi$, we demand it to have a very small relic
density. As we know, the relic density of a thermal relic, e.g.
neutralino LSP, is largely determined by the cross-section of its
annihilation. For $\Psi$, or any strongly interacting particle
like the gluino, there are two main regimes of annihilation:
perturbative era before QCD phase transtion, and non-perturbative
era after QCD phase transition. For the perturbative stage, the
pair annihilation is obviously the inverse process of pair
production, which we discussed in detail in section 4.2. So again,
we can take advantage of the similarity between the gluino and
$\Psi$ and refer to the results for gluino for an estimation. The
upshot is, for a TeV $\Psi$, if we naively only take the
perturbative annihilation into consideration, the relic abundance
$\Omega_ph^2$ is around $10^{-3}$\cite{cheung, pierce, nima},
which is obviously too large if we take seriously the constraints
given in ref.\cite{smith}. For the non-perturbative stage, we can
expect that the difference between the $\Psi$ and the gluinos is
even more blurred since the physics is taken over by hadronic
dynamics when these microscopic particles have become confined in
color-singlet R-hadrons. So it still makes sense to refer to the
corresponding analysis for the gluinos. Since the picture of the
non-perturbative annihilation during the QCD era is not clear,
most of the related study has large theoretical uncertainties and
is quite model-dependent. In most of the known models, it is still
hard to obtain a small enough relic density to satisfy the
constraints from anomalous nuclei in seawater\cite{cheung, pierce,
nima}. Perhaps the most optimistic result is in ref.\cite{cheung},
where $\Omega_{np}h^2\sim10^{-11}$ for a TeV gluino or $\Psi$,
which is still larger than the seawater constraint by a factor of
$10^{15}$. However, there is still hope in that the
non-perturbative physics issue is not clarified yet, and more
importantly, the connection between the relic density calculation
and the existing limits is unclear as discussed above. On the
other hand, the resulting relic density can be small enough to
satisfy the seawater constraints by virtue of some other
reasonable mechanism such as late time second
inflation\cite{cheung} or a low reheating temperature\cite{nima}.
\section{Conclusion}
\indent

   The existence of matter fields in higher representations is an interesting issue to consider
in our search for new physics beyond the SM. In this paper we
extend the MSSM by chiral adjoint matter which is inspired by
intersecting D-brane models. We construct the Lagrangian of this
effective field theory model which is independent of the details
of the original string model. The renormalization group equations
are given for the study of low energy physics. Based on the RGEs,
we carefully discuss the asymptotic freedom problem and
perturbativity of strong coupling in this model. We find that
although asymptotic freedom is hard to be realized here,
perturbativity can be preserved up to the string scale. In this
way, the existence of chiral adjoint matter in low energy physics
does not undermine the motivation of the model, since the
bottom-line of calculable intersecting D-brane models is the
validity of perturbativity at the string scale. We also study the
phenomenology of this model. Under natural assumptions, the scalar
adjoint decays promptly and therefore can be detected at colliders
and is safe cosmologically. The story of the fermion adjoint can
be a bit complex because of the uncertainty in its lifetime. With
a short lifetime, it can be detected at colliders through a
significant beyond the SM signature: $4t+X$. If it has an
intermediate lifetime, we need to check if it respects the limit
from BBN, which we argue is plausible. If it is long-lived or even
the stable LSP, there are many mechanisms to make it satisfy the
constraints from cosmology.

\section *{Acknowledgements}
\indent

   I gratefully acknowledge J.~Wells for his useful comments and
discussions and for carefully reading the manuscript. I also thank
P.~de Medeiros, P.~Kumar,  D.~Morrissey, J.~Qian, J.~Shao,
B.~Thomas, M.~Toharia and D.~Vaman for related discussions, and I
further appreciate D.~Morrissey and B.~Thomas for their helpful
readings of the manuscript.


\end{document}